\documentclass[aps,prd,reprint,superscriptaddress,preprintnumbers,nofootinbib]{revtex4-1}

\usepackage{amsmath}
\usepackage{amsfonts} 
\usepackage{amssymb}
\usepackage{graphicx}
\usepackage[dvipsnames]{xcolor}
\usepackage{hyperref}
\usepackage{tensor}
\usepackage{siunitx}
\usepackage{physics}
\usepackage{appendix}
\usepackage{cancel}

\hypersetup{colorlinks=true,
			urlcolor=purple,
			citecolor=blue,
			linkcolor=magenta,}

\newcommand{\beq}{\begin{equation}}
\newcommand{\eeq}{\end{equation}}

\renewcommand{\]}{\right]}
\renewcommand{\(}{\left(}
\renewcommand{\)}{\right)}

\newcommand{\be}{\begin{eqnarray}}
\newcommand{\ee}{\end{eqnarray}}
\newcommand{\bea}{\begin{eqnarray}}
\newcommand{\eea}{\end{eqnarray}}
\newcommand{\bi}{\begin{itemize}}
\newcommand{\ei}{\end{itemize}}
\newcommand{\ben}{\begin{enumerate}}
\newcommand{\een}{\end{enumerate}}
\def\bes{\begin{equation*}}
\def\ees{\end{equation*}}
\def\bead{\begin{aligned}}
\def\eead{\end{aligned}}
\def\bmat{\left(\begin{matrix}}
\def\emat{\end{matrix}\right)}

\def\cL{{\cal L}}
\def\cO{{\cal O}}

\begin{document}

\title{Heavy fields and the axion quality problem}
\author{Quentin Bonnefoy}
\email{q.bonnefoy@berkeley.edu}
\affiliation{Berkeley Center for Theoretical Physics, University of California, Berkeley, CA 94720, USA}
\affiliation{Theoretical Physics Group, Lawrence Berkeley National Laboratory, Berkeley, CA 94720, USA}

\begin{abstract}
\vspace{0.2cm}
\noindent
The idea that the gravity-induced breaking of global symmetries is encoded in Planck-suppressed operators is not scale-invariant: heavy particles which have nothing to do with the UV completion of gravity can mediate the breaking and produce low-energy operators (partly) suppressed by their own mass scales. Such contributions from heavy fields are typically subdominant with respect to the least Planck-suppressed operators, unless the latter are forbidden, as is usual in solutions to the axion quality problem based on four-dimensional gauge symmetries. Therefore, with a focus on axion physics, I investigate such situations and present toy examples where a non-minimal sector of heavy fields coupling to the axion generates operators whose coefficients are orders of magnitude larger than the naive Planck-suppressed estimates, despite gravity being the only source of breaking. I also stress that the key features of these toy models, namely several families of heavy fields with family-dependent gauge charges, are already present in some non-minimal QCD axion models, using the axiflavon/flaxion and the lighter-than-usual KSVZ axion as examples. This suggests that gauge-symmetry-based solutions to the axion quality problem, or of any quality problem really, need to describe complete UV scenarios or make precise UV assumptions.
\end{abstract}

\maketitle

\section{Introduction}\label{section:intro}

The idea that quantum gravity violates global symmetries \cite{Hawking:1987mz,Giddings:1988cx,Banks:2010zn} is widely accepted, and constitutes a recurrent constraint on particle physics model building: any global symmetry should be accidental, i.e. the possible sources of its breaking should be automatically suppressed as a result of the field and gauge symmetry content of the model. For QCD axions \cite{Peccei:1977hh,Peccei:1977ur,Weinberg:1977ma,Wilczek:1977pj}, which relax the $\bar\theta$ parameter of the Standard Model (SM) to zero (see \cite{Hook:2018dlk,DiLuzio:2020wdo} for recent reviews), the restricting power of this statement is especially clear \cite{Georgi:1981pu,Lazarides:1985bj,Kamionkowski:1992mf,Holman:1992us,Barr:1992qq,Ghigna:1992iv}: even a slight breaking of the Peccei-Quinn (PQ) symmetry can be detected given the sharp bound on strong CP violation, coming for instance from EDM measurements \cite{Baluni:1978rf,Crewther:1979pi,Pospelov:1999mv,Pendlebury:2015lrz,Abel:2020pzs},
\beq
\abs{\bar\theta}\lesssim 10^{-10} \ .
\eeq
Indeed, given a contribution to the axion potential
\beq
\delta V(a)=c\Lambda_a^4\cos\(\frac{na}{f_a}+\beta\)
\label{shiftPotential}
\eeq
(where $\Lambda_a^4\approx \Lambda_\text{QCD}^3m_u$, $f_a$ is the axion decay constant, $n$ an arbitrary integer, and $\beta$ and $c$ arbitrary numbers), the PQ solution to the strong CP problem is spoiled unless
\beq
\abs{c\sin\(\beta-n\bar\theta\)}\lesssim 10^{-10} \ .
\label{qualityRelation}
\eeq
Similar relations arise for light axion-like particles (ALPs), when one imposes that \eqref{shiftPotential} be a minor correction to the potential \cite{Banerjee:2022wzk}. Henceforth, I use the words ``PQ symmetry" to refer to the shift symmetry of any ALP.

Assuming a weakly-coupled UV completion where the ALP arises from the phase of a complex scalar field $\phi$, dubbed ``PQ scalar" below, at its vacuum expectation value (vev), $\phi=\langle\frac{f_a}{\sqrt 2}\rangle e^{ia/f_a}$, \eqref{shiftPotential} originates from operators of the form
\beq
c_\text{UV}\frac{\phi^n}{\Lambda^{n-4}}+{\rm h.c.} \ ,
\label{UVshiftPotential}
\eeq
where $\Lambda$ is a scale of new physics and $c_\text{UV}\equiv c\frac{\Lambda_a^4\Lambda^{n-4}}{(f_a/\sqrt 2)^n}e^{i\beta}$. For values of $f_a$ accounting for astrophysical bounds on simplest QCD axion models \cite{AxionLimits}, the axion quality problem then arises because \eqref{qualityRelation} constitutes a strong tuning on the parameters, even in the fortunate case where no physics other than gravity breaks the PQ symmetry, i.e. when one can take $\Lambda=M_P$ in \eqref{UVshiftPotential}. (By gravity, I really mean any Planck-scale dynamics directly associated to the UV completion of gravity.)

Although unjustified in the formulation above, the tuning of \eqref{qualityRelation} becomes natural when extra ingredients are introduced. For instance, if gauge symmetries (which are respected by gravity) enforce $n\gg 1$, one gets $c\ll 1$ for a fixed, $\cO(1)$ $c_\text{UV}$. A gauged $\mathbb{Z}_n$ symmetry, under which $\phi\to e^{2\pi i/n}\phi$ with $n=\cO(10)$, is the archetypal example \cite{Krauss:1988zc}. Such a mechanism is at the core of gauge-symmetry-based solutions to the axion quality problem, where the PQ symmetry is preserved until operators of very high mass dimensions are included, as an accidental consequence of the gauge structure and field content of the model. Therefore, it is customary to introduce a gauge symmetry in the UV completion of the axion EFT (or better, to make appropriate use of an existing gauge symmetry).

In this note, I explore a simple aspect of this treatment which (to my knowledge) has not been made explicit much in the literature: one implicitly assumes that, besides gravity, no dynamics in the full UV completion generates operators of the form \eqref{UVshiftPotential}. If that does not hold, not all scales in the denominator of \eqref{UVshiftPotential} ought to be $M_P$, even when gravity is the only source of PQ breaking. For instance, the interaction $\phi^3\Phi$ of a complex scalar field $\Phi$ with the PQ scalar respects the PQ symmetry, while gravity could generate the PQ-breaking operator $\phi^{n-3}\Phi^\dagger$ for some integer $n$, which we can take to be suppressed by $M_P^{n-6}$. Integrating out $\Phi$ at tree level, one finds, up to order one Lagrangian parameters,
\beq
\frac{\phi^n}{M_P^{n-6}\Lambda_h^2} +{\rm h.c.}\ ,
\label{worseScaling1}
\eeq
where $\Lambda_h$ is the mass of $\Phi$, thereby increasing the tuning of $c_\text{UV}$ by a ratio $M_P^2/\Lambda_h^2$, or that of $n$ by one unit if $f_a\lesssim \Lambda_h\sim 10^{10}$ GeV. This simplest example shows that postulating that all PQ-breaking operators scale with inverse powers of the Planck mass up to $\cO(1)$ coefficients is a scale-dependent statement. This is consistent with the fact that a single power of $M_P$ in the denominator of \eqref{UVshiftPotential} suffices for gravity to be the only source of PQ-breaking. Note that gauge-symmetry-based solutions to the quality problem do not affect that discussion: for instance, \eqref{worseScaling1} and the $\phi-\Phi$ couplings are invariant under the aforementioned gauged $\mathbb{Z}_n$ symmetry. However, the axion quality problem is strengthened: given a target such as \eqref{qualityRelation}, the order of an appropriate gauged $\mathbb{Z}_n$ symmetry is increased. In more complicated models for the heavy fields, one can even find cases where a few factors of $M_P$ appear in \eqref{worseScaling1}, independently of the value of $n$. Also, the tuning in \eqref{worseScaling1} is sharper than in \eqref{UVshiftPotential} even when $\Lambda_h\gg f_a$, as long as $\Lambda_h\ll M_P$. Consequently, solutions to the axion quality problem need to refer to all high scales, instead of being implemented only at the scale ($\sim f_a$) where the ALP EFT is UV-completed into a linear theory.

The sensitivity of the axion quality problem to UV physics has been noted previously, for instance with respect to heavy QCD axions and UV sources of CP violation \cite{Dine:1986bg,Kitano:2021fdl,Bedi:2022qrd,Valenti:2022tsc,Dine:2022mjw,Zhang:2022ykd}. In the present context, one can argue that the usual evaluation of the tuning persists as long as no new heavy fields are introduced, hence that one is entitled to focus on the one source of PQ breaking which we know for sure, gravity. Here, I choose the (arguably) conservative viewpoint, namely I assume that any heavy field could exist (within observational and consistency limits) and consider models where the problem mentioned above arises. This approach is motivated by the fact that heavy fields are already present in any UV-complete QCD axion model, and are likely to be present in any ALP model to address puzzles of the SM and cosmology.

In what follows, I first discuss general aspects of the mediation of gravity-induced PQ breaking by heavy fields, then I present tree- and loop-level toy examples which generalize and worsen the scaling of \eqref{worseScaling1}. I finally emphasize that some non-minimal QCD axion models already include characteristic features of those toy models, namely several families of heavy fields with family-dependent gauge charges. Arguably, the generic and quantitative form that gravity-induced breaking of global symmetries takes is not known (see \cite{Palti:2019pca,vanBeest:2021lhn} and references therein for recent progress driven by the swampland program). Throughout the paper, I work with four-dimensional field theory models and assume that gravity-induced PQ breaking can be captured by operators such as \eqref{UVshiftPotential}: this assumes that there exists a scale at which the PQ symmetry is linearly realized and that the PQ-breaking effects take the form of local effective field theory (EFT) operators with unsuppressed coefficients (beyond the $M_P$ factors). This does not describe non-local breaking in higher-dimensional models\footnote{On the other hand, the present discussion applies when the breaking proceeds via local operators on 4D branes where the PQ symmetry is global (see e.g. \cite{Izawa:2002qk,Cox:2019rro,Yamada:2021uze}).} \cite{Cheng:2001ys}, where the PQ symmetry can emerge from a higher-dimensional gauge field \cite{Arkani-Hamed:2003xts,Arkani-Hamed:2003wrq,Choi:2003wr,Flacke:2006ad,Grzadkowski:2007xm} as is ubiquitous in the string axiverse \cite{Svrcek:2006yi,Arvanitaki:2009fg}, nor generic non-perturbative corrections from sources associated with the UV completion of gravity \cite{Coleman:1988tj,Gilbert:1989nq,Rey:1989mg,Svrcek:2006yi,Arvanitaki:2009fg,Alonso:2017avz,Hebecker:2018ofv,Alvey:2020nyh,Heidenreich:2020pkc,Cheong:2022ikv}. I also do not include all possible gravity-induced EFT operators, but only sufficiently-illustrative subsets, ignoring possible accidental cancellations. Furthermore, I often use a single PQ scalar $\phi$ (with a PQ charge conventionally normalized to $1$) and a gauge $\mathbb{Z}_n$ to describe the UV completion of the ALP EFT and the solution to the quality problem, but the conclusions are more general. For instance, one can consider an extra $U(1)_X$ gauge group in the UV, and two charged fields $\phi_{1,2}$ whose dynamics has an accidental PQ symmetry, which is realized when their $X$-charges are coprime numbers $p,-q$ such that $p+q$ is large \cite{Barr:1992qq}. Then, an equivalent of the scalar example above is found when introducing couplings $\phi_1^q\Phi$ and $\phi_2^p\Phi^\dagger$, leading to a gauge-invariant PQ-breaking potential $\phi_1^q\phi_2^p/\(M_P^{p+q-6}\Lambda_h^2\)$. Drawing from this example, unless explicitly specified, $\phi^n$ should be interpreted as the lowest-dimensional gauge-invariant PQ-breaking operator (and $n$ as its mass dimension), independently of the details of the PQ sector, of the true nature of the gauge symmetry (which could be discrete \cite{Dias:2002hz,Babu:2002ic,Carpenter:2009zs,Harigaya:2013vja,Dias:2014osa,Harigaya:2015soa,Bjorkeroth:2017tsz,Nakai:2021nyf,Choi:2022fha}, continuous abelian \cite{Babu:1992cu,Hill:2002kq,Fukuda:2017ylt,Duerr:2017amf,Fukuda:2018oco,Bonnefoy:2018ibr,Ibe:2018hir,Suematsu:2018hbu,Bonnefoy:2019lsn} or continuous non-abelian \cite{Cheung:2010hk,DiLuzio:2017tjx,Lee:2018yak,Ardu:2020qmo,Darme:2021cxx,Contino:2021ayn,Darme:2022uzl}) and even of whether the ALP EFT is UV-completed by a weakly or strongly coupled theory \cite{Randall:1992ut,Dobrescu:1996jp,Redi:2016esr,Lillard:2017cwx,Gavela:2018paw,Vecchi:2021shj}.

\section{Mediated PQ breaking and collective effects}

As explained in the introduction, I focus here on the interplay between gravity-induced PQ-breaking operators and the existence of heavy fields $\Psi_h$ at a scale $\Lambda_h\leq M_P$ which mediate this breaking. By gravity-induced PQ breaking, I mean that the PQ symmetry is exact when $M_P\to \infty$. Of course, the heavy dynamics (such as new fields or UV instantons) could break the PQ symmetry in flat space already, in which case one generically expects $\Lambda\to \Lambda_h$ in \eqref{UVshiftPotential}, recovering a similar tuning as for gravity when $f_a$ is not dramatically smaller than $\Lambda_h$. However, regular field theory respects global symmetries (when non-anomalous), so, given an ALP model in the IR or at an intermediate scale and known sources of PQ breaking at these scales, it is consistent (albeit optimistic) to focus on gravity as the only additional source of PQ breaking in the deep UV. Nevertheless, heavy fields remain relevant, for their couplings could mediate gravity-induced breaking. Schematically, instead of
\beq
\cL(\phi,\text{gravity})\supset\cL_\text{PQ}(\phi)+\frac{f_a}{M_P}\cL_{\cancel{\text{PQ}}}(\phi,M_P)
\eeq
I consider
\beq
\bead
\cL_\text{UV}(\phi,\Psi_h,\text{gravity})\xrightarrow[]{E\ll \Lambda_h}\ &\cL_\text{PQ}(\phi,\Lambda_h)\\
&+\frac{f_a}{M_P}\cL_{\cancel{\text{PQ}}}(\phi,M_P,\Lambda_h) \ .
\eead
\eeq
Due to the presence of an extra scale $\Lambda_h$, it is natural to expect (and this indeed happens, as we saw above) that $\cL_{\cancel{\text{PQ}}}(\phi,M_P,\Lambda_h)$ is less suppressed than \eqref{UVshiftPotential} with $\Lambda=M_P$, even when gauge symmetries are introduced.

An interesting aspect of the breaking mediated by fields coupling to a PQ scalar is that it is generically collective: several couplings conspire to break the PQ symmetry, although they individually respect it\footnote{An exception corresponds to the case of real fields which cannot be assigned a charge under the PQ symmetry: an example would be a coupling $\phi^pS$, for $S$ a real scalar, which breaks the PQ symmetry on its own.}. In the simplest complex scalar example of the introduction, $\phi^3\Phi$ (with arbitrary Wilson coefficient $c_3$) respects the PQ symmetry for $q_\Phi=-3$, but so does $\phi^{n-3}\Phi^\dagger$ (with arbitrary Wilson coefficient $c_{n-3}$) for $q_\Phi=n-3$: only when both couplings are present is the PQ symmetry broken. In particular, any PQ-breaking effect is proportional to $c_3c_{n-3}$. 

The idea of collective symmetry breaking, relevant for instance in studies of CP violation \cite{Jarlskog:1985ht,Jarlskog:1985cw} or little Higgs models \cite{Arkani-Hamed:2001nha,Arkani-Hamed:2002sdy,Arkani-Hamed:2002ikv}, usually explains why symmetry breaking is weak: several small factors naturally yield a much smaller overall expression. However, here, it results in an increase of the strength of gravity-induced PQ breaking (for a fixed power $n$ in \eqref{UVshiftPotential}), for the simple reason that several couplings imply several heavy field lines in Feynman diagrams, hence several factors of $\Lambda_h$ (and proportionally fewer factors of $M_P$) in the denominator of \eqref{UVshiftPotential}. In particular, unlike the case of CPV or little Higgs models, other channels contributing to PQ breaking with a lower ``degree of collectiveness" (namely, with fewer couplings conspiring to break the PQ symmetry) contribute less and are unimportant. One can state the same in terms of spurions: several lagrangian terms of low mass dimensions have Wilson coefficients which can be combined to form a spurion of high PQ charge and low inverse mass dimension. Instead, a low degree of collectiveness correlates high PQ charges and high inverse mass dimensions.

Some comments are required in order to clarify this claim. First, I insisted on a fixed power $n$ in \eqref{UVshiftPotential}, since I have in mind that a gauge-symmetry-based solution to the quality problem is implemented. Without this assumption, the operators suppressed by a single Planck mass are the most dangerous, independently of the contributions of hypothetical heavy fields, and a lower degree of collectiveness implies fewer couplings in a Feynman diagram, hence generally fewer external $\phi$ fields (thus fewer $M_P$ factors) or a smaller loop order. When an appropriate gauge symmetry is introduced, it forces any set of PQ-breaking couplings of arbitrary cardinality to generate operators with at least $n$ $\phi$'s. This also explains a second aspect of the claim: fewer couplings mean that each of them multiplies (on average) higher powers of $\phi$, hence one of them comes with a high $M_P$ suppression. This is due to the fact that, by assumption, at least one coupling in the PQ-breaking set should be generated by gravity\footnote{\label{allHeavyFieldsFootnote}A previous statement, namely that a single power of $M_P$ is needed for that, is not in contradiction with the claim: other scales signal even heavier fields, and since we study here the effect of any heavy field, we should as well include those new ones in the discussion and iterate.}. Finally, I should stress that I assume that products of couplings do not drastically reduce the strength of the associated collective effects, or at least that they do not compensate the associated large $M_P/\Lambda_h$ ratios.

\section{$M_P$ counting}

In later sections, I discuss explicit examples of the impact that heavy fields may have on the PQ-breaking potential. They will realize a generic counting of powers of $M_P$ which I present here (see also \cite{Contino:2021ayn}, where the same counting was discussed in order to emphasize that collective PQ breaking is relevant in composite axion models).

First, one should notice that a non-trivial degree of collectiveness is required in order to beat the fully gravity-induced potential (i.e., \eqref{UVshiftPotential} with $\Lambda=M_P$) \cite{Dobrescu:1996jp}, dubbed ``gravity potential" for short below. As I argued above, the simplest collective effect involves two operators of schematic form
\beq
\frac{\phi^k\cO}{M_P^{k+[\cO]-4}} \text{ and } \frac{\phi^{n-k}\cO^\dagger}{M_P^{n-k+[\cO]-4}} \ ,
\label{lowestCollectiveDegree}
\eeq
where $k$ is an arbitrary non-negative integer and $\cO$ is an arbitrary operator of mass dimension $[\cO]$. Only when taken together do these couplings break the PQ symmetry, hence at least one insertion of each is required to generate a PQ-breaking potential, which is therefore suppressed by (at least) $M_P^{n-4+2\([\cO]-2\)}$, independently of the precise graph structure or loop order. The suppression is comparable or larger than that of the gravity potential ($M_P^{n-4}$), except in the aforementioned case of the complex scalar at tree level ($[\cO]=1$). In cases where $k+[\cO]\leq 4$ (or $n-k+[\cO]\leq 4$ if $n-k<k$), the first (second) operator in \eqref{lowestCollectiveDegree} is renormalizable and has no reason to be related to the Planck scale. In that case, the power $k+[\cO]-4$ ($n-k+[\cO]-4$) is negative and the size of the operator is overestimated, nevertheless that does not suffice to overcome the gravity potential. For non-renormalizable operators, not all scales ought to be the Planck scale, but I assume that all heavy fields are resolved (see footnote \ref{allHeavyFieldsFootnote}).

However, this conclusion does not survive complications of the heavy sector: a higher degree of collectiveness brings down the minimal suppression to 
\beq
M_P^{n-4+\sum_i\([\cO_i]-4\)+4} \ ,
\label{highestCollectiveDegree}
\eeq
where $\cO_i$ refer to the operators inserted, among which, given a precise choice for the PQ charges, only one needs to break the PQ symmetry. The missing scales in order to saturate the right mass dimension ($n-4$) must therefore be $\cO(\Lambda_h)$, and the more insertions with $[\cO_i]<4$ the larger the induced potential. Below, I present weakly-coupled examples which realize this, but the same philosophy is applicable to composite axions \cite{Contino:2021ayn}. In that case, $\phi^k$ and $\cO$ should be interpreted as some composite operators which respectively do and do not interpolate with the axion, $k+[\cO]-4$ as the dimension of the associated PQ-breaking spurion and $\Lambda_h$ simply as the confining scale. The latter coincides with the PQ scale, and the composite operators which do not interpolate with the axion play the role of the heavy fields. Fields heavier than the confining scale can be integrated out in the weakly-coupled regime, which is captured by the present analysis.

\section{Toy examples}

Toy examples of \eqref{highestCollectiveDegree} are built by following a simple approach: I couple the different operators $\cO_i$ to $\phi$ and to themselves via renormalizable terms, so that their PQ charges are increasingly large, from $k_1$ to $k_\text{max}$. Then, $n-k_\text{max}$ can be parametrically smaller than $n$ (in the examples below, $k_\text{max}$ scales linearly with the number of heavy fields), yielding PQ-breaking spurions of high charge but low $M_P$ suppression. As shown below, this can be done at tree or loop level.

\subsection{Tree level}

The scaling in \eqref{highestCollectiveDegree} suggests that we should couple $\phi$ to operators of mass dimension $\leq 3$. The extreme case is again realized by scalar fields at tree level. Thus, consider $m$ SM-singlet complex scalar fields $S_i$ of similar masses $m_i\sim\Lambda_h$ and lagrangian
\beq
\sum_{i=1}^{m}m_i^2\abs{S_i}^2+\sum_{i=1}^{m-1}\(\lambda_i\phi^x S_i^\dagger S_{i+1}+{\rm h.c.}\)+\tilde\lambda_1 \phi^2S_1 \ ,
\label{treeToyExample}
\eeq
where $x=1$ or $2$, and I assume that $\tilde\lambda_1\sim m_i\sim\Lambda_h$($\sim\lambda_i$ when $x=1$). This preserves the PQ symmetry under which $S_i$ has charge $x(1-i)-2$. In addition, gravity is expected to generate (at least) the following operator,
\beq
\lambda_{\cancel{\text{PQ}}}\frac{\phi^{n-[2+x(m-1)]}S_m^\dagger}{M_P^{n-x(m-1)-5}}+{\rm h.c.} \ .
\label{manyPhisZnBreakingTree}
\eeq
Integrating out the $S_i$, one obtains a PQ-breaking potential at tree level,
\beq
\frac{\phi^n}{M_P^{n-x(m-1)-5}\Lambda_h^{x(m-1)+1}} \ ,
\label{worseScaling2}
\eeq
up to a factor $c_\text{UV}=\lambda_{\cancel{\text{PQ}}}\frac{\tilde\lambda_1}{\Lambda_h}\prod_i^{m-1}\frac{\lambda_i}{\Lambda_h^x}\prod_i^m\frac{\Lambda_h^2}{m_i^2}$, which may be small for perturbation theory to be a good approximation, while not being as small as $\(\Lambda_h/M_P\)^{x(m-1)+1}$. Demanding that \eqref{manyPhisZnBreakingTree} describes a non-renormalizable operator suppressed by the Planck scale imposes that $x(m-1)\leq n-6$, which, when saturated, leads to a single Planck mass in \eqref{worseScaling2}. In such a case, if $f_a$ is close to $\Lambda_h$, the value of $n$ matters way less to the axion quality problem: one finds that $c_\text{UV}\(\frac{f_a}{10^{10}\text{ GeV}}\)^5\(\frac{f_a}{\Lambda_h}\)^{n-5}\lesssim 10^{-46}$.

Note that \eqref{worseScaling2} is the result of collective breaking: all couplings $\lambda_i,\tilde\lambda_1,\lambda_{\cancel{\text{PQ}}}$ need to be non-vanishing to break the PQ symmetry and yield this result. One may object that the structure of \eqref{treeToyExample} is tuned; it does not even correspond to the most general lagrangian compatible with the PQ symmetry. However, most additional operators, such as $\phi^{n-2}S_1$, only contribute to the PQ-breaking potential at fixed $n$ with larger $M_P$ or loop suppression factors. As said above, focussing on a fixed given $n$ can be imposed by the gauging of a $\mathbb{Z}_n$ subgroup of the PQ symmetry of \eqref{treeToyExample}, which leaves the above story unchanged.

I should stress again that the discussion generalizes to PQ symmetries protected by other gauge symmetries than their gauged $\mathbb{Z}_n$ subgroups. For instance, using the $U(1)_X$ symmetry presented in the introduction, one can choose that $p<q$ even, $x=2,m=q/2$, give a $U(1)_X$ charge $-2ip$ to $S_i$ and replace $\phi\to\phi_1$ in \eqref{treeToyExample}, as well as $\phi^{n-[2+x(m-1)]}\to \phi_2^p$ in \eqref{manyPhisZnBreakingTree}. This has the effect of reducing the Planck suppression of \eqref{UVshiftPotential} from $p+q-4$ to $p-3$. The gauge symmetry could also be non-abelian. As an explicit example, in the construction of \cite{DiLuzio:2017tjx} one introduces a new $SU(N)\times SU(N)$ gauge group factor, as well as a PQ scalar $Y_{a\bar b}$ in the $(N,\bar N)$ representation, such that the leading PQ-breaking operator is $\det Y$ (of dimension $N$). In such a case, a field $S$ of $\mathbb{Z}_n$-charge $-k$ above translates into a field $S_{\bar a_1...\bar a_k,b_1...b_k}$, where all indices are anti-symmetrized, and the gauge indices in the equivalent of \eqref{manyPhisZnBreakingTree} are contracted with epsilon tensors, while those in the equivalent of \eqref{treeToyExample} are contracted with $\delta$'s. In particular, the increasingly large $\mathbb{Z}_n$ charges translate into increasingly large representations.

\subsection{Loop level}\label{loopToyExampleSection}

I now turn to loop-level examples built along the same lines. I focus here on fermions coupled to $\phi$, in order to connect to the following sections, but very similar scalar examples can also be found.

A pair of fermions has mass dimension $3$ and is therefore suitable to alleviate the Planck scale suppression, as seen in \eqref{highestCollectiveDegree}. Thus, consider $m$ Dirac fermions $\psi_i$ of similar masses $m_i\sim\Lambda_h$ and identical vectorlike SM charges, with a lagrangian
\beq
\sum_{i=1}^{m}m_i\bar \psi_i\psi_i+\sum_{i=1}^{m-1}\(\lambda_i\phi \bar \psi_{i,L}\psi_{i+1,R}+{\rm h.c.}\) \ .
\label{loopToyExample}
\eeq
It preserves the PQ symmetry under which $\psi$ has charge $i_0-i$ for some number $i_0$. In addition, gravity is expected to generate the following operator, 
\beq
\lambda_{\cancel{\text{PQ}}}\frac{\phi^{n-(m-1)}\bar \psi_{m,L} \psi_{1,R}}{M_P^{n-m}}+{\rm h.c.} \ .
\label{manyPhisZnBreakingLoop}
\eeq
In the limit where $m_i=\Lambda_h$ $\forall i$, the one-loop Coleman-Weinberg potential induced by the $\psi_i$ has a compact expression and contains an operator\footnote{The fact that gauge-invariant PQ-breaking potential terms hide within fermion mass matrices has already been noted in \cite{Cheung:2010hk}: in this section, we are basically exploring how small the scale suppressing these terms can be.}
\beq
\frac{\phi^n}{16\pi^2M_P^{n-m}\Lambda_h^{m-4}} \ ,
\label{worseScaling3}
\eeq
times a factor $16\pi^2c_\text{UV}=\frac{2(-1)^m}{(m-1)(m-2)}\lambda_{\cancel{\text{PQ}}}\prod_{i=1}^{m-1}\lambda_i$. Again, choosing $m=n-1$ is compatible with the effective nature of \eqref{manyPhisZnBreakingLoop} and makes \eqref{worseScaling3} suppressed by a single power of $M_P$ (the $16\pi^2$ factor clearly does not compensate the large $M_P/\Lambda_h$ factors). The other remarks made in the tree-level case apply here as well. Notice that the model is vectorlike, hence the fermions would not introduce any gauge anomaly, were a $\mathbb{Z}_n$ subgroup of the PQ symmetry gauged or $\phi$ charged under an additional $U(1)$ gauge symmetry. Another consequence of the fact that the fields have SM vectorlike charges is that the only couplings from \eqref{loopToyExample} which are linear in the axion field and violate the PQ symmetry are those which are real and flavor-diagonal in mass basis. In a more general flavor basis, they are captured by the following flavor invariants \cite{Bonnefoy:2022rik},
\beq
I^{(k=0,...,m-1)}\equiv\Re\Tr\(\[MM^\dagger\]^kM\tilde M^\dagger\) \ ,
\label{PQbreakingInvariants}
\eeq
where $m$ is as above the number of flavors, $M$ is the mass matrix and $\tilde M$ is the matrix of Yukawa couplings to the axion (both are $m\times m$ complex matrices). Via the Coleman-Weinberg potential, those invariants, associated to PQ-breaking couplings linear in the axion field, generate a tadpole for the axion, and reciprocally: only those couplings can generate a tadpole. Consequently, the tadpole of the PQ-breaking potential must be captured by a combination of the invariants in \eqref{PQbreakingInvariants}. This can be explicitly checked in the model of \eqref{loopToyExample}, upon expanding $\phi=\langle\frac{f_a}{\sqrt 2}\rangle e^{ia/f_a}$,
\beq
\bead
&\qquad I^{(k=0,...,m-2)}=0 \ , \\ 
I^{(m-1)}=-\sqrt 2&\frac{\Lambda_h^mM_P^{n-m}}{f_a}\(\frac{f_a}{\sqrt 2}\)^{n-1}\Im\(\lambda_{\cancel{\text{PQ}}}\prod_{i=1}^{m-1}\lambda_i\) \ .
\eead
\eeq
The imaginary part is indeed the quantity which generates an axion tadpole from the Coleman-Weinberg potential of \eqref{worseScaling3}, and the high degree of the non-vanishing invariant illustrates the collective effect at play: there are $m$(the $m_i$)$+m-1$(the $\lambda_i$)$+1$($\lambda_{\cancel{\text{PQ}}}$)$=2m$ couplings conspiring to break the PQ symmetry and forming $c_\text{UV}$, so only the invariant $I^{(m-1)}$, which is of polynomial degree $2m$, can feel this collective effect. Had we considered a gravity-induced operator $\phi^{n-(j-i)}\bar \psi_{j,L} \psi_{i,R}$ in \eqref{manyPhisZnBreakingLoop}, we would have witnessed contributions from all invariants of degree $\geq 2(\abs{j-i}+1)$ (with a stronger $M_P$ suppression).

Sticking with the same category of models, a lower value than $m=n-1$ can be sufficient to obtain a single $M_P$ suppression, upon using a chiral model. Inspired by the Froggatt-Nielsen models of \cite{Alonso:2018bcg,Smolkovic:2019jow,Bonnefoy:2019lsn}, let us consider the same field content as above with the same SM charges, but with the following lagrangian,
\beq
\sum_{i=1}^{m}y_i\phi^\dagger\bar \psi_{i,L}\psi_{i,R}+\sum_{i=1}^{m-1}\lambda_i\phi \bar \psi_{i,L}\psi_{i+1,R}+{\rm h.c.} \ ,
\label{loopToyExampleChiral}
\eeq
which selects linearly-decreasing chiral PQ charges ($i_0-2i$ for $\psi_{i,L}$ and $i_0-(2i-1)$ for $\psi_{i,R}$). I assume that $\lambda_i\leq y_i$ so that the masses are mostly determined by the diagonal terms and $\Lambda_h\sim y\langle\phi\rangle$. Gravity is now expected to generate the following operator,
\beq
\lambda_{\cancel{\text{PQ}}}\frac{\phi^{n-(2m-1)}\bar \psi_{m,L} \psi_{1,R}}{M_P^{n-2m}}+{\rm h.c.} \ ,
\eeq
leading to a one-loop Coleman-Weinberg potential which contains, in the limit $y_i=y$ $\forall i$,
\beq
\frac{\phi^{n-m+2}}{16\pi^2M_P^{n-2m}\abs{y}^{m-2}\phi^\dagger{}^{m-2}} \ ,
\label{worseScaling4}
\eeq
times a factor $16\pi^2c_\text{UV}=\frac{2(-1)^m}{(m-1)(m-2)}y^*{}^{m}\lambda_{\cancel{\text{PQ}}}\prod_{i=1}^{m-1}\lambda_i$, leaving one (two) Planck mass(es) when $m=(n-1)/2$ for $n$ odd ($m=n/2-1$ for $n$ even). Working at fixed $n$ can again be imposed by an additional gauge symmetry and a non-vanishing charge for $\phi$; notice however that this makes our model chiral and subject to anomaly cancellation constraints. For a $\mathbb{Z}_n$ subgroup of the PQ symmetry, one requires that \cite{Ibanez:1991hv,Preskill:1991kd,Banks:1991xj,Ibanez:1992ji} $T^{G_{SM}}_\psi m(m+1)=0 \mod n$, where $T^{G_{SM}}_\psi\delta^{ab}\equiv\Tr(T_{R(G_{SM})}^aT_{R(G_{SM})}^b)$ for $R(G_{SM})$ the representation of the $\psi_i$ under the factor $G_{SM}=SU(3)_C$ or $SU(2)_L$ of the SM gauge group $SU(3)_C\times SU(2)_L\times U(1)_Y$. This is satisfied for instance when $n=10,m=4$ and $R(G_{SM})$ is the fundamental representation of $SU(3)_C$ or $SU(2)_L$. $U(1)_Y$ charges are free, and $\mathbb{Z}_n^3$ or $\mathbb{Z}_n\times$gravity anomalies can be cancelled by SM-neutral chiral fermions.

\section{QCD axion models}

Any QCD axion model can be coupled to ad-hoc heavy modes with the features of our toy examples, but the resulting models are not particularly motivated and their mere existence may fail to convince the reader (or me) that they seriously threaten the axion quality. However, non-minimal well-motivated QCD axion models often include several heavy fields and, depending on the precise model building, might already contain the ingredients of our toy examples. I illustrate this in the case of axiflavon/flaxion models \cite{Davidson:1981zd,Wilczek:1982rv,Ema:2016ops,Calibbi:2016hwq} and of the lighter-than-usual KSVZ axion model \cite{Hook:2018jle,DiLuzio:2021pxd,DiLuzio:2021gos}. The discussion is more general (see e.g. \cite{Bonnefoy:2018ibr} for an instance in clockwork-like gauge models), but it nevertheless relies on precise features of the models under consideration, on which I comment.

\subsection{Axiflavon/flaxion models}\label{axiflavonSection}

The axiflavon/flaxion scenario \cite{Davidson:1981zd,Wilczek:1982rv,Ema:2016ops,Calibbi:2016hwq} unifies the PQ symmetry with a Froggatt-Nielsen (FN) symmetry \cite{Froggatt:1978nt}, so as to solve the strong CP problem and explain the flavor hierarchies in one go. The FN mechanism postulates a $U(1)$ symmetry in the UV, broken by a small charged spurion $\epsilon$. Upon choosing appropriately the FN charges of the SM chiral fermions, the flavor hierarchies arise from the necessary powers of $\epsilon$ which compensate for the FN charges of the Yukawa couplings. For instance, if the Higgs is FN-neutral (which can always be arranged by composing with a global $U(1)_Y$ transformation), the Yukawa coupling $\bar Q_iu_j\tilde H$ in the up-quark sector comes with a coefficient suppressed by $\epsilon^{\abs{q_{Q_i}-q_{u_j}}}$, where $q_{Q_i}$ ($q_{u_j}$) is the FN-charge of the $i$-th ($j$-th) generation left-handed (right-handed up) quark. In addition, the breaking encoded by $\epsilon$ can be spontaneous, in which case one needs to consider the associated Nambu-Goldstone boson (NGB). In the axiflavon/flaxion scenario, one chooses the FN charges so that the symmetry has an $SU(3)_C$ anomaly and the NGB is the QCD axion.

More precisely, in the minimal scenario, one assumes a spontaneous breaking by the vev of a complex scalar field $\langle\phi\rangle$. Then one write $\epsilon=\abs{\frac{\langle\phi\rangle}{M}}$, for a scale $M$ slightly larger than $\langle\phi\rangle$. In the up-quark Yukawa sector again, the couplings (except that of the top) arise from non-renormalizable operators,
\beq
\tilde y_{u,ij}\(\frac{\phi}{M}\)^{q_{Q_i}-q_{u_j}}\bar Q_iu_j\tilde H \ ,
\label{upQuarkFN}
\eeq
where $\tilde y_{u,ij}=\cO(1)$ and $\tilde y_{u,ij}\epsilon^{q_{Q_i}-q_{u_j}}$ is the usual Yukawa coupling. If $q_{Q_i}-q_{u_j}<0$, one uses $\phi^\dagger$ instead. 

The scale $M$ signals the presence of additional heavy FN modes, whose interactions are structured very similarly to those of \eqref{loopToyExample}. Indeed, one can generate \eqref{upQuarkFN} from
\beq
a_i\bar Q_i\psi_1\tilde H+\eqref{loopToyExample}+b_j\bar\psi_mu_j\phi \ ,
\label{UVcompleteAxiflavon1}
\eeq
provided $m=\abs{q_{Q_i}-q_{u_j}}$, $\prod_{i=1}^mm_i=M^m$, $a_ib_j\prod_{i=1}^{m-1}\lambda_i=\tilde y_{u,ij}$ and the $\psi_i$ have the same SM charges as the right-handed up quarks. In terms of the PQ charges defined in section~\ref{loopToyExampleSection}, this lagrangian fixes those of $Q_i,u_j$ to be $i_0-1,i_0-(m+1)$ respectively. Since the model of section~\ref{loopToyExampleSection} is a sub-sector of the present UV completion of the axiflavon/flaxion model, it is subject to the same limitations when it comes to solving the axion quality problem via gauge symmetries. For instance, one often encounters in the literature $\abs{q_{Q_1}-q_{u_1}}=8$, in which case the loop-induced operator \eqref{worseScaling3} overcomes the gravity one by a factor $(\epsilon M_P/f_a)^4/(16\pi^2)\gg 1$.

The presence of larger-than-expected PQ-breaking operators does not rule out the axiflavon/flaxion scenario, but confirms that a solution to the axion quality problem needs to include (at least) the full dynamics at scale $M$. For instance, focussing on first-generation up quarks and assuming $\abs{q_{Q_1}-q_{u_1}}=8$, one can modify \eqref{UVcompleteAxiflavon1} as follows:
\beq
a_i\bar Q_1\psi_1\phi+\eqref{loopToyExample}\big\vert_{\mathbf{2}}+\tilde\lambda\bar \psi_{4,L}\psi_{5,R}\tilde H+\eqref{loopToyExample}\big\vert_{\mathbf{1}}+b_j\bar\psi_8u_1\phi \ ,
\label{UVcompleteAxiflavon2}
\eeq
where $\eqref{loopToyExample}\big\vert_{\mathbf{2}}$ corresponds to \eqref{loopToyExample} with $\psi_{i=1,...,4}$ in the representation $(\mathbf{3},\mathbf{2},1/6)$ of the SM gauge group, while $\eqref{loopToyExample}\big\vert_{\mathbf{1}}$ refers to \eqref{loopToyExample} with $\psi_{i=5,...,8}$ in the representation $(\mathbf{3},\mathbf{1},2/3)$. Then, the SM gauge invariance restricts the appearance of \eqref{manyPhisZnBreakingLoop} to $\bar\psi_{m,L}\psi_{1,R}\to\bar\psi_{4,L}\psi_{1,R}$ or $\bar\psi_{8,L}\psi_{4,R}$ (at most), which yields a PQ-breaking Coleman-Weinberg potential comparable or smaller than the gravity one. In parallel, gravity can generate
\beq
\lambda_{\cancel{\text{PQ}}}\frac{\phi^{n-6}\bar \psi_{8,L} \psi_{1,R} H}{M_P^{n-6}}+{\rm h.c.} \ ,
\eeq
which leads to a one-loop potential which differs by a factor $\frac{v^2M_P^2}{M^4}$ (where $v$ is the electroweak vev) from the gravity one. Since $M\gtrsim \langle\phi\rangle\gtrsim 10^{10}$ GeV due to bounds on flavor-changing neutral currents, we see that the gravity potential appropriately captures the order of magnitude of the full PQ-breaking potential\footnote{I focus in this paper on the one-loop Coleman-Weinberg potential induced by the heavy fermions. In principle, given the large hierarchies present here, higher-order corrections, coming for instance from four-Fermi operators at two-loops, should be considered, but I do not engage in this discussion and simply stress that model building options exist at one-loop.}. Another option is to stick with \eqref{UVcompleteAxiflavon1} but use the two scalar fields $\phi_{1,2}$ charged under an additional $U(1)_X$, as explained in the introduction. By choosing $p=\cO(1),q\gg1$ (e.g. $p=1,q=10$), and $\abs{q_{Q_i}-q_{u_j}}\leq q$, one can only use $\phi_1$ instead of $\phi$ in \eqref{UVcompleteAxiflavon1}, and the gauge symmetry strictly prevents terms such as \eqref{manyPhisZnBreakingLoop} from being generated (at the level of $\phi_1$ alone, the $U(1)_X$ and PQ symmetries are impossible to disentangle\footnote{In order for the PQ mixed anomalies with the SM not to be aligned with those of $U(1)_X$, some chiral fermions need to couple to $\phi_2$. This could be the case of down quarks, upon choosing that $\abs{q_{Q_i}-q_{d_j}}$ are multiples of $q$. However, such an assignment generically leaves out $U(1)_X$-SM mixed gauge anomalies, which should be cancelled by additional chiral heavy fermions. Since extra matter with SM charges is observationally constrained to be heavy, those additional fields need to obtain a mass from coupling to $\phi_{1,2}$. Therefore, they generically contribute to the PQ anomalies and to the axion couplings to SM gauge bosons, weakening the predictivity of the axiflavon/flaxion setup. Models without any additional fermion (beyond those required by the FN mechanism) can be found by considering chiral FN fields, as sketched at the end of section \ref{loopToyExampleSection} \cite{Bonnefoy:2019lsn}.}). This example highlights the importance of designing explicitly the UV completion: if $p=4,q=7$, the gravity potential is as suppressed as that when $p=1,q=10$, but the former case allows one to write \eqref{manyPhisZnBreakingLoop} with $\phi^{n-(m-1)}\to \phi_2^p$ and a $M_P^{p-1}$ suppression (instead of $M_P^{p+q-4}$).

\subsection{Lighter-than-usual KSVZ axion}

The (to my knowledge, only) paradigm which produces a QCD axion lighter than in simplest models makes use of $n$ copies of the SM and of an axion which realizes non-linearly the $\mathbb{Z}_n$ exchange symmetry between the copies \cite{Hook:2018jle,DiLuzio:2021pxd,DiLuzio:2021gos},
\beq
\mathbb{Z}_n \ : \quad \text{SM}_i \to \text{SM}_{i+1} \ , \ a\to a+\frac{2\pi f_a}{n} \ .
\label{lightAxionZnAction}
\eeq
In order to respect this $\mathbb{Z}_n$ symmetry, all gluons must couple to the axion and have $\theta$ parameters shifted by increasing multiples of $\frac{2\pi}{n}$, which has the effect of reducing the axion mass by a factor $n^{3/2}\(m_u/m_d\)^n$ with respect to the standard value (a mild $1/n$ tuning is also needed to correctly solve the strong CP problem). Several UV completions of this EFT were presented in \cite{DiLuzio:2021pxd}, and I will focus here on the weakly-coupled one. One introduces $n$ copies of KSVZ quarks $\psi_i$ \cite{Kim:1979if,Shifman:1979if} in the fundamental of the strong gauge groups $SU(3)_{C,i}$, coupled to the same PQ scalar $\phi$, so that the $\mathbb{Z}_n$ symmetry, which can be gauged, acts as
\beq
\mathbb{Z}_n \ : \quad \text{SM}_i \to \text{SM}_{i+1} \ , \ \phi\to e^{\frac{2\pi i}{n}}\phi \ , \ \psi_i \to \psi_{i+1} \ ,
\eeq
and the most general lagrangian reads
\beq
\sum_{k=1}^nye^{\frac{2\pi ik}{n}}\phi\bar \psi_{k,L}\psi_{k,R}+{\rm h.c.} \ .
\label{lightAxionPQ}
\eeq
The model enjoys a global PQ symmetry under which $\phi\to e^{i\alpha}\phi,\psi_{k,L/R}\to e^{\pm i\alpha/2}\psi_{k,L/R}$. Unlike above, the gauge symmetry forbids any mixing between different $k$-sectors, but gravity could generate terms like\footnote{I note that a $k$-independent Dirac mass for all $\psi_{k,L/R}$ is allowed by the gauge and $\mathbb{Z}_n$ symmetries, while breaking the PQ symmetry. (Although such a mass term breaks the PQ symmetry in each $k$-sector independently, destructive interference between all sectors is such that the induced Coleman-Weinberg potential is proportional to $\phi^n$, as dictated by the exact $\mathbb{Z}_n$ symmetry.) This term could be induced by physics associated to the UV completion of gravity (such as new instantons), however I only focus here on the impact of $M_P$-suppressed EFT operators.}
\beq
\sum_{p=2}^\infty\sum_{k=1}^n\frac{y_p}{M_P^{p-1}}e^{-\frac{2\pi ikp}{n}}\phi^\dagger{}^p\bar \psi_{k,L}\psi_{k,R}+{\rm h.c.} \ ,
\label{lightAxionPQBreaking}
\eeq
respecting the $\mathbb{Z}_n$ but not the PQ symmetry.

One finds that the leading term in the one-loop Coleman-Weinberg potential is suppressed by a factor $M_P^{(n+2\[n \mod 3\])/3}\abs{\phi}^{2(n-\[n \mod 3\]-6)/3}$, i.e. the power of $M_P$ is roughly reduced by a factor $3$ with respect to the gravity potential, or equivalently the required tuning of $n$ which achieves a satisfying axion quality is worsened by a factor $3$. For illustration, when $n=0\mod 3$ and $n\geq6$, one obtains
\beq
\frac{c_{n/3}\phi^n}{16\pi^2M_P^{n\over 3}\abs{\phi}^{\frac{2(n-6)}{3}}}+{\rm h.c.}
\eeq
where $c_{n/3}=\frac{\(-1\)^{n/3}54}{(n-6)(n-3)}\(yy_2^*\)^{n/3}\abs{y}^{\frac{2(6-n)}{3}}$ (or $-9\(yy_2^*\)^2$ when $n=6$). The power $n/3$ of the Planck mass can be understood by a spurion analysis, as follows. The coefficient of $\phi^p\bar\phi^q\bar \psi_{k,L}\psi_{k,R}$ has a spurious PQ charge $1+q-p$, and is suppressed by $M_P^{p+q-1}$, hence it has a ``charge to $M_P$-power ratio" of $(1+q-p)/(p+q-1)$, which is maximal (and $=3$) when $p=0,q=2$. Therefore, the least suppressed potential is obtained by computing graphs with as many $y_2$ from \eqref{lightAxionPQBreaking} as possible. When $n=1\mod 3$, one insertion of $y_3$ is needed, when $n=2\mod 3$, two insertions of $y_3$ or one of $y_4$ are needed.

In this model, the realization \eqref{lightAxionZnAction} of the $\mathbb{Z}_n$ symmetry is such that gauge multiplets are not its irreducible representations. Nevertheless, in order to more easily connect to the previous sections and pinpoint the collective PQ-breaking effects, it is convenient to use the $\mathbb{Z}_n$ irreducible representations,
\beq
\chi_l\equiv \sum_{k=1}^ne^{-\frac{2\pi i(k-1)l}{n}}\psi_k \ ,
\eeq
which are such that $\chi_l\to e^{2i\pi l/n}\chi_l$ under the action of $\mathbb{Z}_n$. They are not representations of the gauge group, but gauge interactions are irrelevant for the one-loop contribution of the fermions to the potential of the PQ scalar. In terms of those $\chi_l$ fields, \eqref{lightAxionPQ}-\eqref{lightAxionPQBreaking} resemble the patterns of couplings of the previous models,
\beq
\sum_{l=0}^{n-1}\(\tilde y\phi\bar\chi_{l+1,L}\chi_{l,R}+\sum_{p=2}^\infty \frac{\tilde y_p}{M_p^{p-1}}\phi^\dagger{}^p\bar\chi_{l,L}\chi_{l+p,R}\)+{\rm h.c.} \ ,
\eeq
where $\chi_m\equiv \chi_{m\,\text{mod}\,n}, \tilde y \equiv ye^{\frac{2\pi i}{n}}$ and $\tilde y_p \equiv y_pe^{-\frac{2\pi i p}{n}}$. 

As for the axiflavon/flaxion scenario, the present discussion is not threatening the whole paradigm of the lighter-than-usual QCD axion, but should simply be seen as a refined quality constraint when building the UV completion of the axion EFT.

\subsection{Other models}

The two QCD axion models above share the following features: several families of heavy fields, as well as unsuppressed flavor-changing couplings between them. It is then no surprise that (the majority of) models which do not have those features are immune to the effects discussed in this note. With a single family of heavy fields, with the notable exception of the complex scalar, one has too few couplings to accommodate large collective effects. This can be understood from \eqref{highestCollectiveDegree}: a large collective effect requires several operators of dimension $\leq 3$, but those are limited for a given number of fields and a given gauge structure. Moreover, even when there are multiple families, which may be demanded by anomaly cancellation when there are fermions, collective effects can turn out to be heavily Planck-suppressed. In particular, it happens in models where the heavy fields are given family-blind gauge charges, which is for instance realized in the aforementioned models of \cite{Barr:1992qq} and \cite{DiLuzio:2017tjx}. (Note that the family-independent $U(1)_X$ charges crucially differentiate the former model from that of section \ref{loopToyExampleSection}.) In such situations, the renormalizable Yukawa terms can be diagonalized in flavor space while being expressed in terms of irreducible gauge multiplets, and any family-changing lagrangian term, necessary for a collective effect, is as much constrained by the gauge invariance as it would be for a single family. (Terms which would vanish from permutation symmetry arguments with a single family may provide caveats to this claim.) As already anticipated in section \ref{axiflavonSection}, flavor-dependent gauge charges for heavy fermions can be allowed in specific cases, for instance if they do not permit to write flavor-changing Yukawa couplings beyond sets of a few families. Model 1 of \cite{Duerr:2017amf} provides an example, where two families are connected at renormalizable level and the induced Coleman-Weinberg potential is as Planck-suppressed as the gravity potential.

\section{Conclusion}

Accounting for the coupling of particle physics models with global symmetries to gravity is now standard practice, especially when the global symmetry should be free of significant breaking. I focussed on the common assumption that gravity-induced global symmetry breaking manifests itself as non-renormalizable operators suppressed by appropriate powers of the Planck scale. Despite this treatment being well-known, it is rarely emphasized that Planck-suppressed symmetry-breaking operators involving several fields can generate other Planck-suppressed symmetry-breaking operators, once the heaviest fields are integrated out. Such contributions, which are suppressed by the heavy fields masses but also by the Planck scale, are subdominant with respect to the lowest-dimensional Planck-suppressed operators, unless the latter vanish due to the structure of the theory. This precisely happens in solutions to the axion quality problem based on four-dimensional gauge symmetries. In such cases, as I illustrated above, the contributions of heavy fields can easily predominate by several orders of magnitude over the naive estimates based on a full $M_P$ suppression. Although ad-hoc heavy fields can always be postulated (or at least considered, in an agnostic bottom-up perspective), I stressed that they may already be present in non-minimal models of QCD axions, which I illustrated with two representative examples, the axiflavon/flaxion and the lighter-than-usual KSVZ axion. I also argued that the many models with few families of heavy fields, or with family-independent gauge charges, are likely not to be concerned.

Consequently, gauge-symmetry-based solutions to the quality problem need to be implemented at any scale above the PQ scale, and in any sector which can possibly communicate with the axion. More generally, this discussion suggests that studies of the axion quality problem (or really of the fate of any global symmetry coupled to gravity) in terms of Planck-suppressed operators in an effective framework at an intermediate (e.g., PQ) scale should either include explicit UV assumptions (e.g., a ``desert" until the Planck scale or a secluded PQ sector) or use a conservative $M_P$ scaling for the symmetry-breaking operators. If one works instead with a full-fledged UV completion where all heavy fields below $M_P$ are specified, one should nevertheless evaluate all the relevant (classical or quantum) contributions to the PQ-breaking lagrangian.

\section*{Acknowledgments}

I thank Thomas Biek\"otter, Jonathan Kley, Soubhik Kumar, Claudio Andrea Manzari and Pablo Qu\'ilez for discussions. I also thank Thomas Biek\"otter and Jonathan Kley for their comments on the manuscript, and Pablo Qu\'ilez for pointing out the arguments of \cite{Contino:2021ayn}. I am supported by the Office of High Energy Physics of the U.S. Department of Energy under Contract No. DE-AC02-05CH11231.

\bibliographystyle{apsrev4-1_title}
\bibliography{biblio.bib}

\end{document}